\documentclass{elsart}

 \usepackage{graphics}

 \usepackage{graphicx}
 \usepackage{epsfig}

\usepackage{amssymb}

\begin{document}

\begin{frontmatter}




\title{Cellular automaton for bacterial towers}


\author{J. O. Indekeu\corauthref{cor1}}
\ead{joseph.indekeu@fys.kuleuven.ac.be}
\corauth[cor1]{Corresponding author} and
\author{C. V. Giuraniuc}
\ead{claudiu.giuraniuc@fys.kuleuven.ac.be}
\address{Laboratorium voor Vaste-Stoffysica en
Magnetisme,\\Celestijnenlaan 200 D, Katholieke Universiteit
Leuven,\\ B-3001 Leuven, Belgium}

\begin{abstract}
A simulation approach to the stochastic growth of bacterial towers
is presented, in which a non-uniform and finite nutrient supply
essentially determines the emerging structure through elementary
chemotaxis. The method is based on cellular automata and we use
simple, microscopic, local rules for bacterial division in
nutrient-rich surroundings. Stochastic nutrient diffusion, while
not crucial to the dynamics of the total population, is
influential in determining the porosity of the bacterial tower and
the roughness of its surface. As the bacteria run out of food, we
observe an exponentially rapid saturation to a carrying capacity
distribution, similar in many respects to that found in a recently
proposed phenomenological hierarchical population model, which
uses heuristic parameters and macroscopic rules. Complementary to
that phenomenological model, the simulation aims at giving more
microscopic insight into the possible mechanisms for one of the
recently much studied bacterial morphotypes, known as ``towering
biofilm", observed experimentally using confocal laser microscopy.
A simulation suggesting a mechanism for biofilm resistance to
antibiotics is also shown.
\end{abstract}

\end{frontmatter}

\section{\large{Introduction}}
Bacteria can either grow freely or adhere to surfaces. In the latter
case one obtains structures ranging from compact colonies with
self-affine surfaces, over diffusion-limited self-similar
aggregates, to towering-pillar or mushroom shaped biofilms
\cite{Herma}. Bacterial biofilms are commonly present in daily life;
they typically flourish on substrates in aqueous environments, from
sewage pipes to human teeth. In addition to performing physical
measurements, it is necessary to introduce theoretical models for
biofilm development. These models can be used in the prediction
and/or understanding of the system's evolution in experiment or
simulation and they can also improve the understanding of intrinsic
features. As an example, the efficiency of an antibiotic can depend
on the spatial distribution and organization of bacteria.

Two different approaches have been developed over time: they use
either microscopic or macroscopic rules. While macroscopic models
can give more easily a global image of the biofilm, the
microscopic ones describe better the behaviour at the level of the
smallest biofilm unit - the ``cell". Ideally suited as elemental
units of a cellular automaton approach, these cells contain
typically a very small number of bacteria, or lots of nutrient
particles, or simply consist of water-filled compartments.
Biophysical properties are not imposed {\em a priori} but emerge
as a result of local microscopic rules \cite{Herma}.

One of the best known theoretical biology models for population
growth, which has since long been applied in various other fields,
is the Eden Model \cite{Eden}. Various features have been added to
this model, in order to describe different types of situations
that appear in specific cases. Other approaches are diffusion
limited aggregation (DLA)\cite{Wit} and, more generally, the
cellular automata methods \cite{Wol}. In this article we shall
present a method based on this last technique, which amounts to
nutrient-limited aggregation in which chemotaxis is the driving
mechanism and diffusion plays a minor role.

\section{Model description} The general framework we employ
is the cellular automata (CA) method. CA are discrete dynamical
systems whose behavior is completely specified in terms of a local
relation (rules). CA can be thought of as a stylized universe. Space
is represented by a uniform grid, with each site or ``cell"
containing a few bits of data. Time advances in discrete steps, and
the laws of the universe are expressed by a simple recipe - say, a
small look-up table - through which at each step each cell computes
its new state from that of its close neighbors. The system's laws
are local and uniform.

For our purpose the environment is a three-dimensional simple
cubic lattice of size ($l\times w\times h = 81\times 81 \times
27$). Each lattice site is in one of a finite number of distinct
states. Usually the number of states is small but, in principle,
any finite CA model over a finite alphabet can be defined. The
state of a cell (site) changes according to some rules. There are
two distinct classes of automata: synchronous, which allows for
simultaneous update of all cells' states, and asynchronous. The
updating order can be deterministic or random in the asynchronous
class. The local rules of a certain CA are essentially determined
by the considered neighborhood, consisting of the 6 nearest
neighbors, 12 next-nearest and 8 next-next-nearest neighbors of a
given site in the cubic lattice, 26 in total. We have used
asynchronous updating.

We start with a homogeneous substrate layer of bacteria of size
$81\times 81 \times 10$ submerged in water. The water initially
occupies a $81 \times 81 \times 17$ volume. We deposit at random
large nutrient chunks on the substrate, conveniently approximated in
shape by compact rectangular pillars of cells, of size $27\times
27\times 10$. The pillars are deposited with probability $P$ over a
grid of 9 macroscopic square substrate plaquettes of area $27 \times
27$, each containing many substrate cells. This process creates a
strongly non-uniform nutrient supply partly covering the substrate.
Bacterial cells may multiply {\em provided} they eat neighboring
nutrient cells (nutrient is then replaced with bacteria). This
process is characterized by a growth probability $G$ and represents
binary division coupled to elementary chemotaxis, as the bacteria
grow and spread preferentially in the direction of highest nutrient
concentration. Clearly, the greater the number of neighboring cells
occupied by nutrient, the greater is the probability that a
bacterial cell will divide. For the time being, we disallow
diffusion, and leave an examination of nutrient diffusion effects to
a later section.

Technically, we choose at random a pair of neighboring cells and if
the pair consists of nutrient and bacteria we replace the nutrient
by bacteria with probability $G$ and then pass on to another
randomly chosen pair. We take measurements at time steps of
($l\times w\times h$) updates, so that each cell has on average been
visited once in every time step, and record bacterial population and
nutrient configuration.

\section{Results}
\subsection{Nutrient-limited growth without diffusion}
We assume a probability $P$ for putting at random nutrient pillars
over the coarse-grained substrate grid and a probability $G$ for
bacterial cells to consume adjacent nutrient cells and subsequently
divide, thereby occupying also the nutrient cell. As can be seen in
the vertical section shown in Figure 1, the bacteria develop towers
in the places where the nutrient was deposited at random. Note the
developing roughness of the originally flat bacteria/nutrient
interface. {\em Within a tower} the growth process is Eden-like (in
three dimensions), since all perimeter cells carry nutrient.
Eventually, however, the bacteria fill up the towers and perfectly
flat walls and rooftops result.

\begin{figure}[htbp]
\centerline{\epsfxsize=8cm \epsfbox{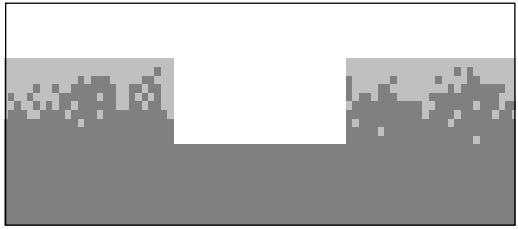}} \caption{Vertical
section ($81 \times 27$) through the cellular lattice system of
bacteria (dark), nutrient (light) and water (white), after 20 time
steps. The nutrient deposition probability $P$ equals 0.33 and the
bacterial growth probability $G$ equals 0.2. Two bacterial towers
grow where nutrient pillars were placed. Eden-like growth takes
place inside the pillars until eventually they are filled
compactly with bacteria. Note that bacterial cells which seem
detached from the colony in this 2D section, are connected in the
3D structure.}
\end{figure}

Let us denote by $N(t)$ the bacterial population, in units of cells,
{\em in excess} of the initially present substrate population, so
that $N(0)=0$. Plots of $N(t)$, measured every time step, reveal a
characteristic {\em sigmoid} shape, common for various biological
growth processes approaching a finite carrying capacity \cite{Mur}.
Since the deposition of nutrient is stochastic, the carrying
capacity $K$ depends on the sample, and consequently a
$K$-distribution arises, each $K$ being simply equivalent to the
total mass of deposited nutrient pillars. The speed of saturation of
the population towards $K$ is, of course, faster when $G$ is bigger
and for each population curve we can discern three distinct regimes
as illustrated in Figure 2, for which we took $P=0.33$.

\begin{figure}[htbp]
\centerline{\epsfxsize=8cm  \epsfbox{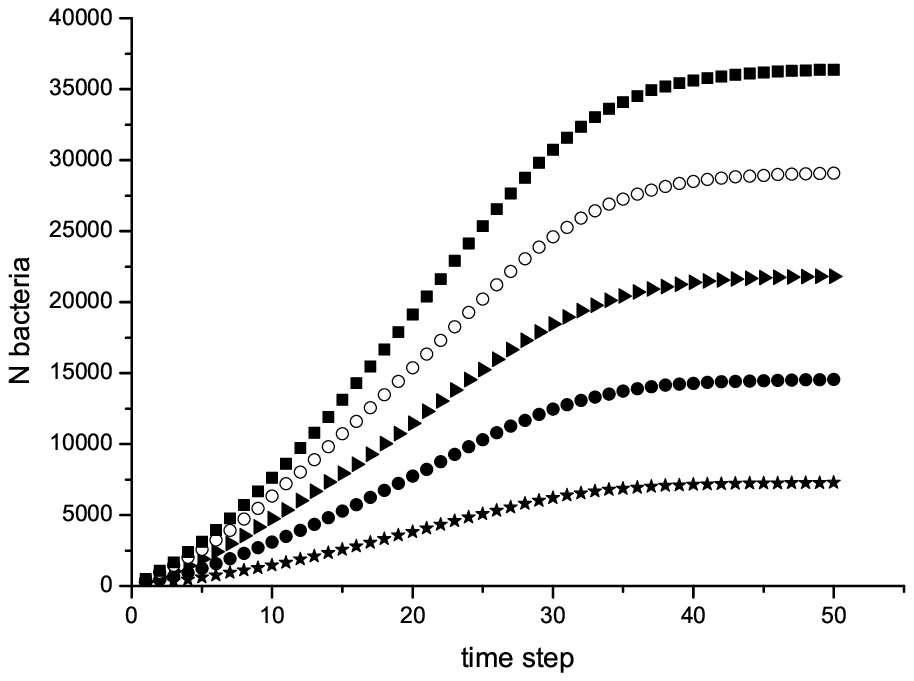}\epsfxsize=8cm
\epsfbox{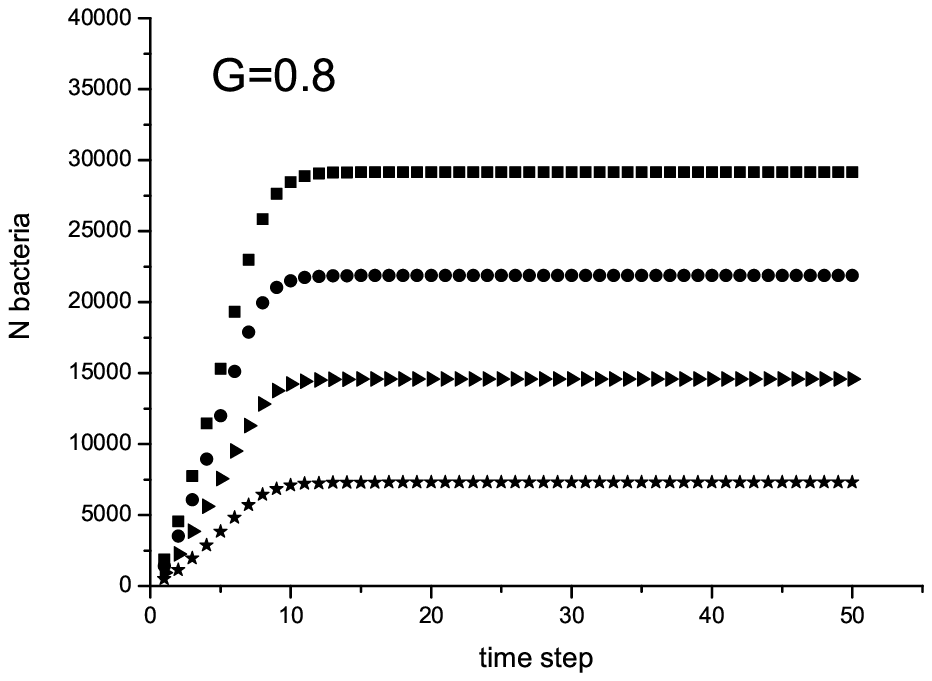}} \caption{The number of bacterial cells versus
time for growth probability $G=0.2$ (left) and $G=0.8$ (right).
The different curves, each with its own carrying capacity
$N(\infty) = K$, correspond to different numbers of randomly
deposited nutrient pillars. The fluctuations around a given curve
are small, so that each curve accurately represents a narrow band
of very similar curves. For $G=0.2$ the linear saturation time
$\tau_{lin,sat}$ equals about 65 time steps, for all curves, and
for $G=0.8$ about 12 time steps, for all curves. The growth of
$N(t)$ first accelerates and then comes to a halt, leading to a
characteristic sigmoid shape.}
\end{figure}

i) The initial growth is not exponential - it would be exponential
in free binary division - but is limited by the amount of nutrient
in the first horizontal layer adjacent to and in contact with the
bacterial substrate. The population increase is just proportional
to the {\em contact area} $A$ and to $G$, leading to a {\em
linear} initial growth. This stage can be characterized by a {\em
linear saturation time} $\tau_{lin,sat}$, proportional to $1/G$.
This is the average time needed to build a bacterial tower in the
absence of interface fluctuations (cf. ``mean-field"
approximation), and is found by simple linear extrapolation of the
initial growth. ii) The growth soon accelerates as the area of the
bacteria/nutrient contact boundary increases by {\em interface
roughening} and the appearance of nutrient ``holes" (see Fig.1).
The roughness, caused by height fluctuations, is limited by the
finite horizontal width of the nutrient pillar(s), resulting in a
saturating contact area, again implying a linear growth regime.
iii) This regime in turn crosses over to an exponentially fast
convergence of the growth towards $K$, which is the asymptotic
behaviour for long times. Indeed, the bacteria reaching the
ceiling of the nutrient pillar fall short of nutrient and
eventually all nutrient cells are consumed. The sigmoid shape
arises from the succession of these three regimes i)-iii). The
dependence of the bacterial population on time can be understood
most clearly from the time evolution of the area $A(t)$ of the
bacteria/nutrient contact interface, shown in Figure 3.

\begin{figure}[htbp]
\centerline{\epsfxsize=8cm \epsfbox{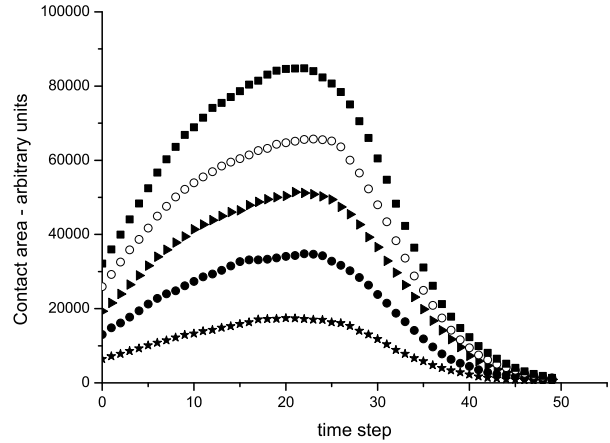}} \caption{Contact area
of the bacteria/nutrient interface in units of pairs of neighboring
cells consisting of adjacent bacteria and nutrient. The nutrient
deposition probability is $P = 0.33$ and the growth probability
$G=0.2$. The curves shown are proportional to the time derivative of
the population curves in Fig.2 (left). Note the finite constant for
initial times, the maximum for intermediate times and the
exponential decay for long times.}
\end{figure}

Actually, we have
\begin{equation}
\frac{d N(t)}{dt} \propto G A(t)
\end{equation}

Asymptotically for long times, the decrease of the nutrient mass,
$M(t)$, which equals the increase of the bacterial population
$N(t)$, is simply proportional to the number of nutrient cells,
since these become isolated, whence independent, and surrounded by
neighboring bacterial cells. Thus we have
\begin{equation}
 d M \propto - M  G\, dt
\end{equation}
It follows that the asymptotic behavior of $N$ for long times can be
written as
\begin{equation}
N(t) \sim K (1 - e ^{- (t-t_0)/\tau}),
\end{equation}
where the saturation time $\tau$ is proportional to $1/G$ and
depends only weakly on the probability $P$ of depositing nutrient
pillars, because these can to a first approximation (especially
for small $P$) be considered as ``non-interacting". Numerically,
$1/\tau \approx 0.85 \pm 0.1$ for $G=0.8$, implying a very short
time scale for the actual exponential saturation in Fig.2 (right)
compared to the linear saturation time. The time $t_0$ denotes the
moment for which a zero excess population, $N(t_0)\equiv 0$, would
result by extrapolating backwards the asymptotic behavior. This
time equals about $t_0 = 16$ for all curves with $G = 0.2$ and
about $t_0 = 5$ for all curves with $G = 0.8$ (Fig.2). It is
remarkable that the three characteristic times $\tau_{lin,sat}$,
$\tau$ and $t_0$ are, within the accuracy of our computations,
independent of the carrying capacity $K$.

\subsection{Nutrient-limited growth with nutrient diffusion}

We have performed simulations similar to the ones described above,
but in which we have included stochastic nutrient diffusion. A
probability $I$ is introduced for the {\em interchange} of adjacent
cells, one occupied with nutrient and the other with water. This
model is closer to physical, or biological, reality since the
nutrient in general is soluble in water. In our simulations
diffusion turns out to be necessary for generating the observed
porous morphologies with fractal-like surface roughness and possible
overhangs leading to mushroom-shaped towers, also seen
experimentally \cite{Herma}. Figures 4 and 5 show relevant vertical
and horizontal sections through the developing biofilm structure.

\begin{figure}[htbp]
\centerline{\epsfxsize=8cm \epsfbox{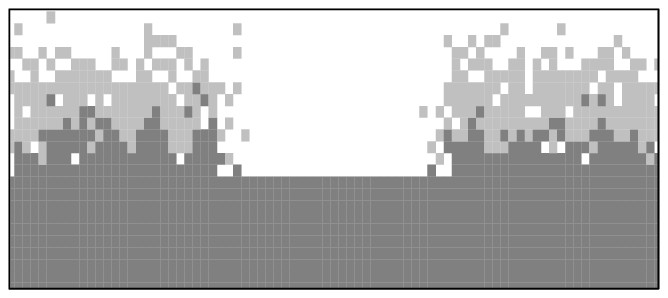}\epsfxsize=8cm
\epsfbox{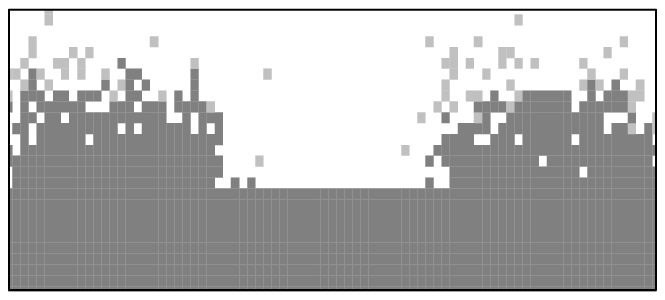}} \caption{Vertical sections through the
biofilm at an early time $t=6$ (left) and at a later time $t=20$
(right), for $P=0.33$, $G=0.6$ and diffusion probability $I=0.6$.
A cross-over from Eden to DLA growth occurs by the time most of
the nutrient pillars is consumed.}
\end{figure}

In the presence of nutrient diffusion, bacterial reproduction
generates a two-step biofilm growth. Firstly, Eden-like growth
driven by chemotaxis takes place in the nutrient pillars (as for
zero diffusion). Secondly, DLA-type of growth roughens and ramifies
the bacterial towers as the wandering nutrient recombines by
aggregation. Clearly, the long time DLA process determines the
biofilm morphotype. A remarkable property of the fractal character
of the emerging structure becomes detectable in Fig.5 (horizontal
slices through the biofilm). Sections through the top layers reveal
a cascade or hierarchy of length scales. Water ``lakes" in bacterial
``land" develop into bacterial ``islands" surrounded by water, over
a height difference of just a few layers. This is strikingly similar
to the structure that is inherent in the hierarchical population
model, where it is imposed {\em ad hoc}. In contrast, here in the
simulations it occurs spontaneously.

\begin{figure}[htbp]
\centerline{\epsfxsize=8cm \epsfbox{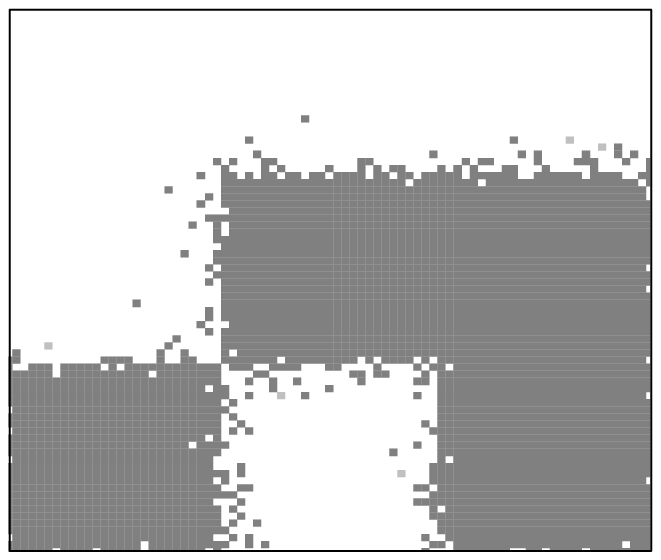}\epsfxsize=8cm
\epsfbox{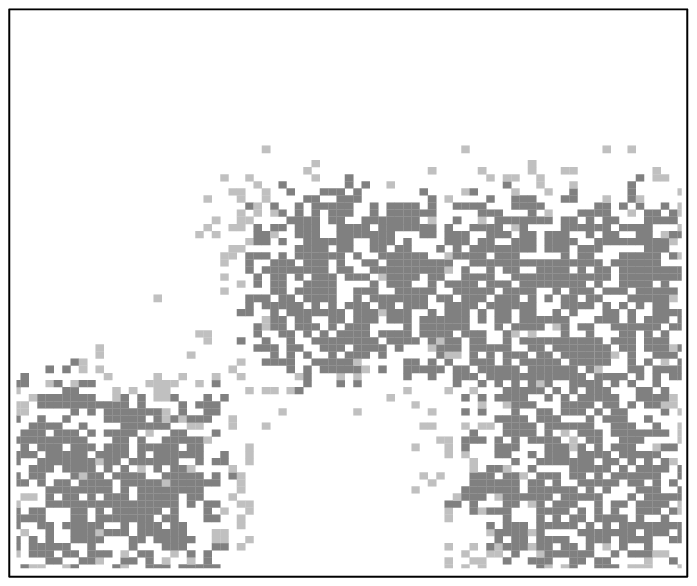}} \centerline{\epsfxsize=8cm
\epsfbox{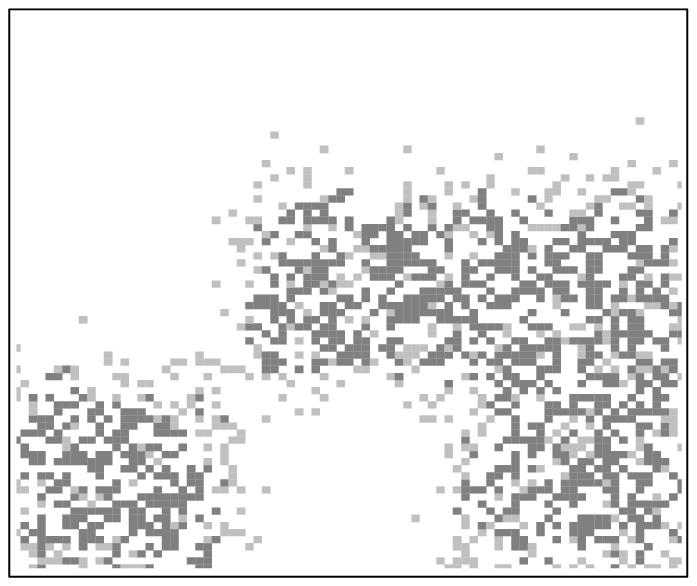}\epsfxsize=8cm \epsfbox{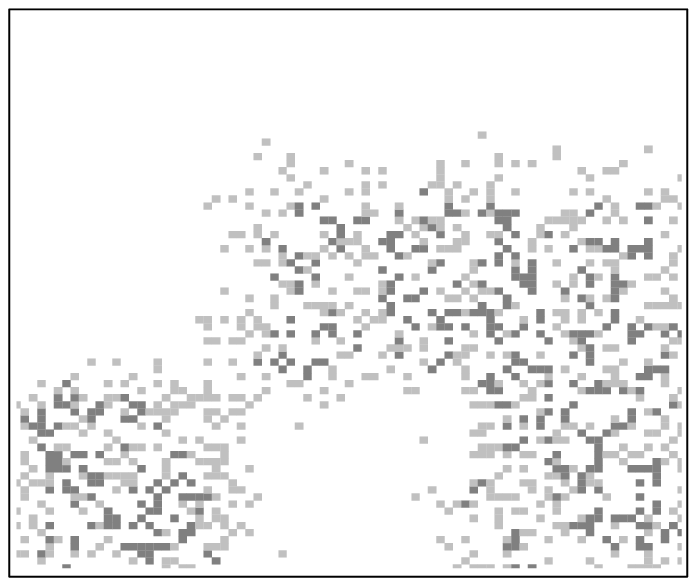}}
\caption{Horizontal sections through the biofilm, for the same set
of probabilities as in Fig.4 and at time $t=20$ (cf. Fig.4 - right).
In the upper left a section just above the substrate (layer 11)
shows the compact base of the four bacterial towers and some
aggregation on the periphery. The remaining sections, at layer 19
(top right), 20 (bottom left) and 21 (bottom right), show the
fractal character of the tower tops. A hierarchy of bacterial
islands (dark) is observable, amidst nutrient (light) and water
(white).}
\end{figure}

In Figure 6 we observe once more the basic sigmoid shape of the
$N(t)$-curves describing the time evolution of the bacterial
population, for 10 samples. Thus the shape of the population curves
is robust with respect to the new perturbation; it is not much
influenced by diffusion, at least not for short and medium times.
The new feature brought about by diffusion is the existence of a
{\em long tail} in $N(t)$, beginning after the formation of
bacterial towers and ending when all detached nutrient has diffused
``back" towards, and is eaten by, the bacterial colonies. This tail
is still exponential, but with a decay time roughly 100 times longer
than in the absence of diffusion. For example, for $P=0.33$ and
$G=0.8$, we find $1/\tau = 0.009 \pm 0.001$. Interestingly, this
saturation rate turns out to be independent of $I$ provided $I>0$,
at least to the accuracy of our computations so far (averaging over
samples, and for $I=0.2, 0.4, 0.6, 0.8$ and 1). One may speculate
that this could have been anticipated qualitatively, since weak
diffusion leads to minor detachment of nutrient and swift
recombination. On the other hand, strong diffusion entails more
detachment but also greater mobility, whence also swift aggregation.
The DLA process' efficiency seems largely independent of the
diffusion probability.

\begin{figure}[htbp]
\centerline{\epsfxsize=8cm \epsfbox{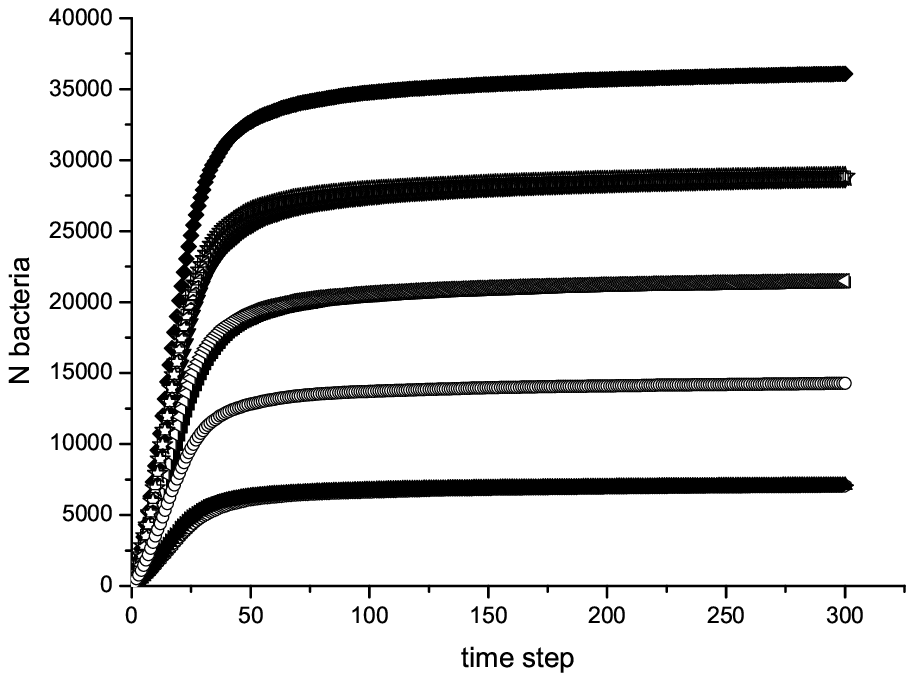}\epsfxsize=8cm
\epsfbox{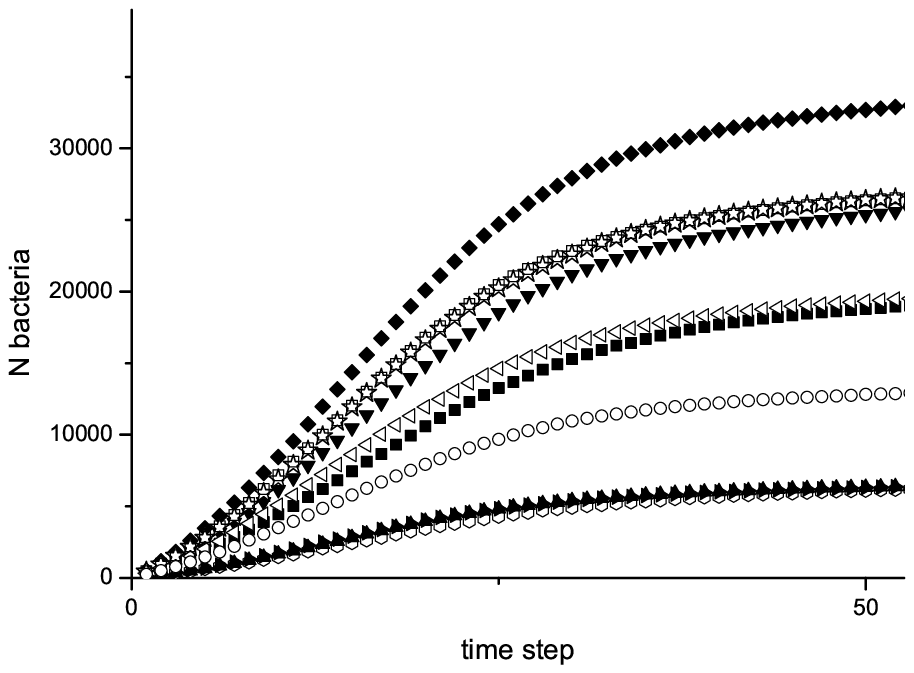}} \caption{The number of bacterial cells versus
time for growth probability $G=0.2$ and diffusion probability
$I=0.2$, for 10 samples randomly drawn with $P=0.33$. The
different curves, each with its own carrying capacity $N(\infty) =
K$, correspond to different numbers of randomly deposited nutrient
pillars (each pillar eventually leads to 7290 bacterial ``cells").
In the presence of diffusion the fluctuations of $N(t)$, given a
fixed number of nutrient pillars and thus given $K$, are not
small, so curves bunch into bands rather than into a sharp line.
On the left the full evolution is shown on a time scale long
enough to observe saturation, while on the right the detail for a
shorter time scale is presented, just long enough to form the
towers.}
\end{figure}

The exponential character of the saturation can be understood as
follows. For long times, long after the bacterial towers are
formed, the remaining water-dispersed nutrient particles can be
considered as independent walkers - an assumption verified by
keeping track of the asymptotic {\em monomer} character of
nutrient matter in the simulations - and thus the asymptotics is
again governed by $ d M \propto - M dt$, implying exponential
decay.

Just like in the previous simulations without diffusion, a
carrying capacity distribution is found, which is controlled by
the initial stochastic deposition of nutrient pillars. For a given
number of nutrient pillars, the carrying capacity $K$ is fixed. To
illustrate how this $K$-distribution emerges we present a
histogram (Fig.7) giving the frequency of occurrence of $N(10)$,
the population after ten time steps, for 1000 samples with
$P=0.33$, $G=0.8$ and $I=0.8$. Note the fine structure of the main
peaks. To optimize the resolution we have excluded from the
viewing window the trivial peak at $N=0$ (no nutrient was put) and
a small but not negligible eighth peak (outside of view, to the
right). Note that, as time progresses, these peaks sharpen and
converge to a simple set of ten $K$-values corresponding to a
population filling all nutrient cells originally contained in the
pillars. This sharpening is particular to our simple model, and
will no longer occur when additional realistic ingredients such as
antibiotics, detachment of bacteria and rinsing water flow is
allowed, causing fluctuations of $N(\infty)$.

\begin{figure}[htbp]
\centerline{\epsfxsize=8cm \epsfbox{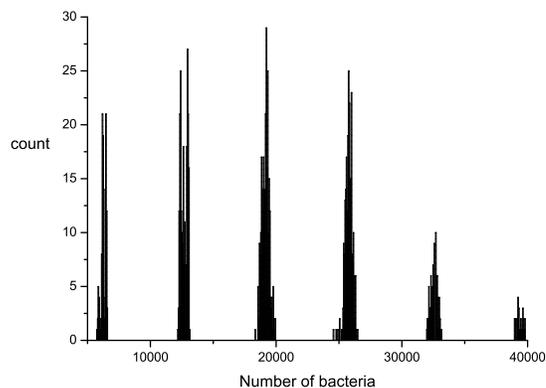}} \caption{Population
distribution over 1000 samples taken at time step $t= 10$, for
$P=0.33$, $G=0.8$ and $I=0.8$, reflecting the ``density-of-states"
of the bands of bunching $N(t)$-curves at that particular time.}
\end{figure}

\section{Comparison with the hierarchical population model}

The random hierarchical population model \cite{Ind} generates
bacterial towers through a more macroscopic construction algorithm.
A typical {\em local} population $N(x,y; t)$ grown over a substrate
in the $xy$-plane, for $t=4$ (after four generations) and for
$P=0.33$ is shown in Fig.8.

\begin{figure}[htbp]
\centerline{\epsfxsize=8cm \epsfbox{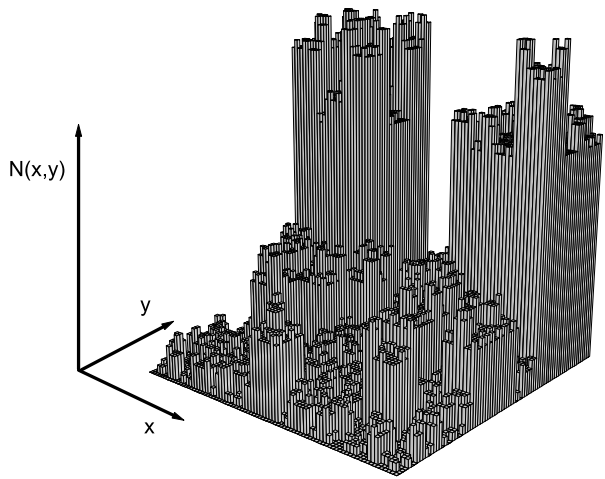}} \caption{Bacterial
population density from the hierarchical model, consisting of two
``towers" and similar replicas on smaller length scales. Note that
not height but density $N$ is plotted, which is equivalent to a
compactification or compression of the real-space
three-dimensional biofilm against the substrate.}
\end{figure}

A vertical (left) and a horizontal (right) section through the
density $N(x,y)$ of Fig.8 are shown in Fig.9. Obviously, bacterial
colonies of the type depicted in Fig.4 (right) and in Fig.5 are
fairly well represented by Figs.8-9. The differences reside in the
details of the self-similar properties intrinsic to the asymptotic
long-time behavior of the hierarchical model and not to that of the
simulations.

\begin{figure}[htbp]
\centerline{\epsfxsize=8cm \epsfbox{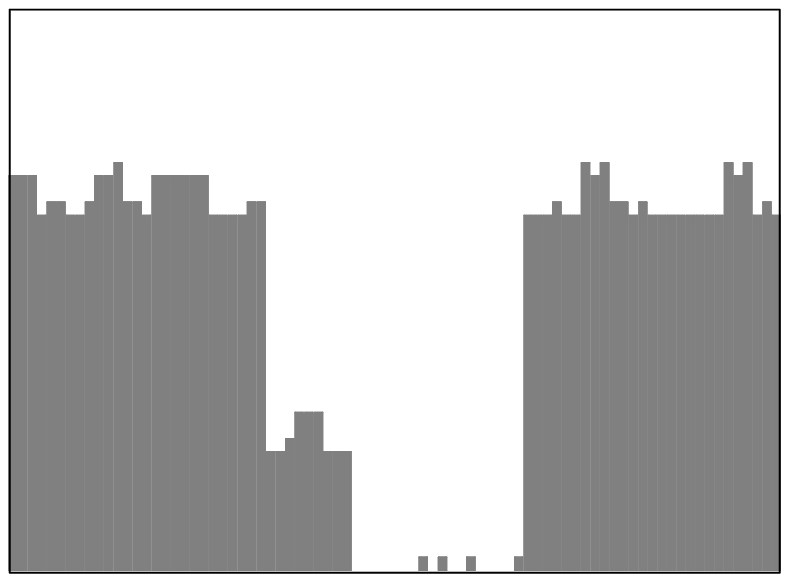}\epsfxsize=8cm
\epsfbox{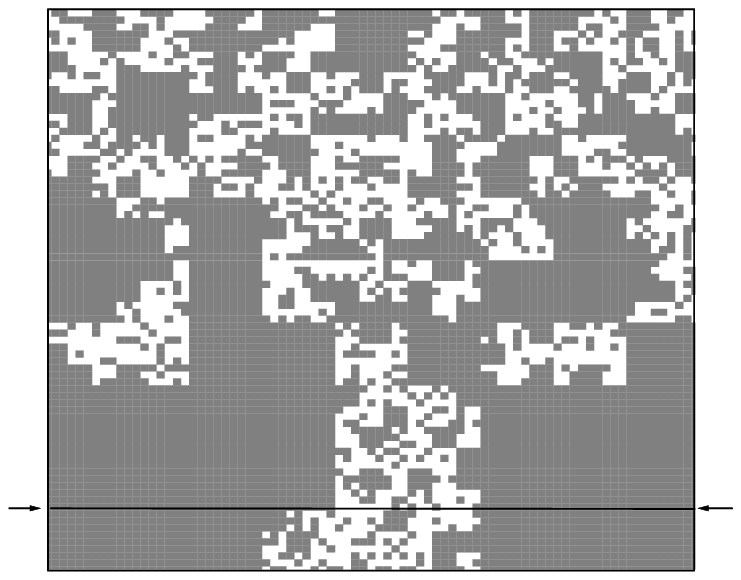}} \caption{Vertical (left) and horizontal
(right) sections through the density plot of Fig.8. The horizontal
section is taken at the bottom of the density ``landscape". The
vertical section is taken perpendicular to the line indicated in
the horizontal section. The figures are rotated by $180^{\circ}$
in the $xy$-plane relative to Fig.8.}
\end{figure}

Typical population curves $N(t)$ produced in the hierarchical
model are displayed in Fig.10, for $P=0.33$ and length rescaling
parameter $\lambda = 3$. After a short transient, the populations
saturate exponentially rapidly to their carrying capacity. For
long times, self-similar carrying capacity distributions result
\cite{Ind}, which bear a resemblance to that of Fig.7, except in
the details of the fractal geometry. We now turn to the important
question of whether we can provide a microscopic basis for the
rescaling factor $\lambda$.

\begin{figure}[htbp]
\centerline{\epsfxsize=8cm \epsfbox{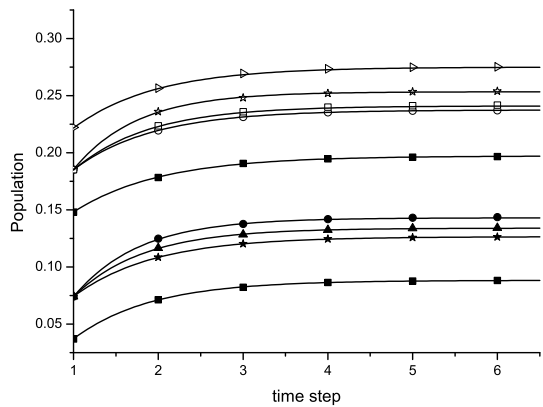}}
\caption{Hierarchical-model population curves for 9 samples
generated with $P=0.33$. Fast exponential convergence to the
respective carrying capacities is observed, the time scale of
which is set through a heuristic length rescaling factor $\lambda$
(here, $\lambda = 3$). The population corresponds to the integral
over the density $N(x,y)$ defined over the unit square, and the
time step is in units of generations. The curves through the data
are guides to the eye.}
\end{figure}

In the exactly soluble random hierarchical population model the
average excess population of bacteria $\bar N(t)$ obeys the simple
time evolution in discrete time $t=n$ \cite{Ind},
\begin{equation}
\bar N(n) = (1-\lambda^{-n})\bar K,
\end{equation}
where $\bar K$ denotes the quenched average carrying capacity.
This quantity is proportional to the birth probability $B$ (in the
present context the probability of death, $D$, equals zero,
because no antibiotic or other bactericidal agent is
considered)\cite{Ind}. Therefore, $\bar K \propto B$. The
exponential saturation is characterized by a rescaling factor
$\lambda$, which is introduced {\em ad hoc} in that model to
incorporate various possible conditions leading to a convergent
population (limited nutrient, temperature boost or quench,
diffusion without convection, bacterial motility, etc.). The
asymptotic behavior of $\bar N$ is conspicuous by rewriting (4) in
the form
\begin{equation}
\bar N(t) = (1-e^{-(\log \lambda) t}) \bar K
\end{equation}

In our CA simulations without diffusion, the towers fill up
completely and compactly with bacteria (eventually without surface
roughening), and the average carrying capacity is proportional to
$P$, the nutrient pillar deposition probability. Precisely the
same asymptotic result can be achieved in the hierarchical
population model, provided we set $B=P$ in the first generation,
and $B=1$ in all subsequent generations, while restricting the
growth to the plaquettes where towers were grown in generation 1
(i.e., setting $B=0$ for substrate areas where no nutrient was
put). Clearly, the correspondence between $\lambda$ in that model
and the variables of our present simulation is
\begin{equation}
\log \lambda = 1/\tau \propto G
\end{equation}
This convergence rate applies not only to the average but also to
each curve of Fig.10 individually, since the process is
self-averaging for long times [6]. From (6) we learn that a
constant microscopic (``cellular") growth probability $G$ combined
with a finite nutrient supply, fixing $K$, leads to an exponential
saturation, justifying the use of and providing a microscopic
interpretation for the constant rescaling factor $\lambda$ in the
hierarchical population model.

Including nutrient diffusion in the simulations affects the
asymptotic long-time behavior in that a non-zero nutrient/water
interchange probability $I$ strongly increases the saturation
time. Further, rough biofilm surfaces result. The hierarchical
model is ideally suited to generate similar roughness, with
marginal fractal geometry, in the spatial distribution of the
population density $N(x,y;t)$ (Fig.8). Note that this density does
not provide full information about the three-dimensional biofilm
structure - in contrast with the simulations which give the full
3D bacterial ``cell" population $N(x,y,z;t)$ and nutrient ``cell"
mass $M(x,y,z;t)$. The hierarchical-model population density
yields a compactification in the form of a single-valued density
or height profile defined over a 2D support, the substrate. Thus,
Fig.8 shows a relevant population landscape from the hierarchical
model, conform with the present 3D simulations including
diffusion. Once again, the length rescaling factor $\lambda$ can
be interpreted using the more microscopic parameters $G$ and $I$
of the CA simulations. Our simulations so far suggest that (6)
remains valid for $I \neq 0$.

\section{Conclusions and perspectives}

We have proposed a microscopic cellular automaton describing
bacterial tower development on a planar substrate in a strongly
non-uniform nutrient environment. The ``cells" contain bacteria,
nutrient or water. A growth probability $G$ governs the stochastic
chemotactic consumption of nutrient cells by adjacent bacterial
cells, causing the former to be replaced by bacteria. A
probability $I$ controls the interchange of neighboring nutrient
and water cells, provoking nutrient diffusion. Nutrient is
deposited only once, at the beginning, in the form of drops or
pillars, with probability $P$. The strongly non-uniform nutrient
supply causes bacterial colonies to grow into towers, and the
absence of nutrient refuelling induces exponential saturation of
the bacterial population towards a carrying capacity $K$ which
depends on $P$. In view of the quenched random deposition a
$K$-distribution is found, rather than a single final population.

Typical sigmoid curves are obtained for the time-dependence of the
bacterial population. The asymptotic behavior for long times can
be mapped onto similar saturation behavior displayed in the
hierarchical population model. The phenomenological length
rescaling factor in that model, necessary to render a convergent
population, can be justified at a more microscopic level through
its correspondence to $G$ (and $I$) found in the present work.
This finding lends support to the applicability of that model as a
phenomenological tool for generating fractal towering-pillar
biofilm structure.

The simulations can be improved easily by adding new features
while keeping maximal simplicity. We consider especially the
possibility to add some antibiotic, deposited with probability
$Q$, which is capable of killing bacterial cells (with extinction
probability $E$), and to look for its bactericidal efficiency. In
this process an adjacent antibiotic/bacteria pair of cells gets
replaced by a pair water/dead-cell. Further, bacterial motility
can be included through an additional water/bacteria cell
interchange probability, allowing bacterial cells to {\em detach}.
We have implemented these ideas in a simulation in which on each
plaquette nutrient is deposited with $P= 0.33$, antibiotics is
deposited with $Q=0.33$ and nothing is put with probability
$1-(P+Q)$. In a first approximation only {\em dead cells} are
allowed to detach. All diffusive interchange probabilities
(water/nutrient, water/dead-cell, and water/antibiotic) are set to
$I=0.6$. Further, $G=0.8$ and $E=1$. Fig.11 shows a horizontal
section through the resulting biofilm structure after about 20
time steps. Interestingly, a bacterial tower has arisen in a
nutrient-rich region (right), but only a relatively small hole
resulted in the neighbouring antibiotic-rich area (middle). The
mechanism which prevents massive extinction is the formation of a
layer of dead cells mixed with water, separating the lower-lying
bacteria from the bactericidal agent. This feature has no
counterpart in the hierarchical population model, which is
symmetric with respect to interchange of ``birth" and ``death",
but may be instrumental in biofilm resistance to antibiotics.

\begin{figure}[htbp]
\centerline{\epsfxsize=14cm \epsfbox{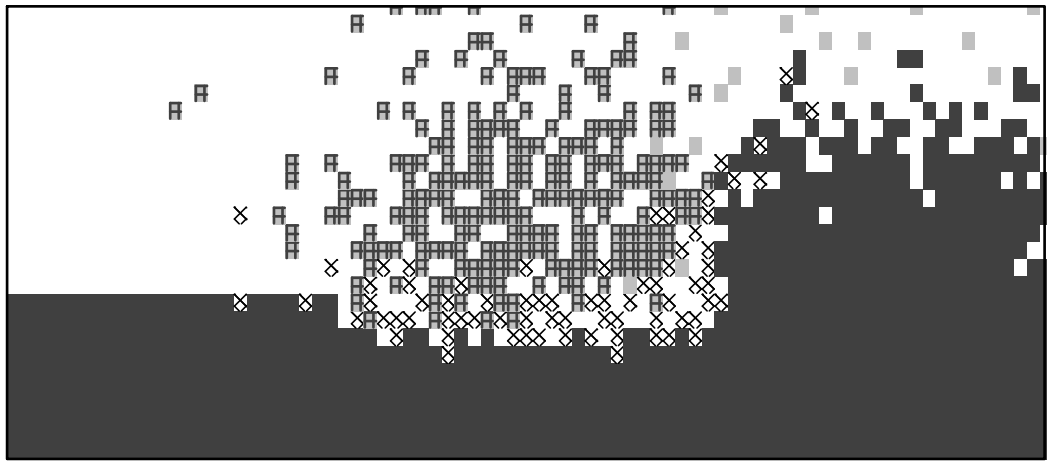}} \caption{Vertical
section through a simulated 3D ``towering-pillar biofilm"
consisting of water cells (white), bacterial cells (black), dead
bacterial cells (x), nutrient cells (grey) and antibiotic
cells(+). Diffusion is allowed for nutrient, antibiotic and dead
cells.}
\end{figure}

In a next step, ``rinsing" can be added, entailing removal of
detached bacteria, antibiotics and/or nutrient by water flow.
Bacterial division in low-nutrient surroundings can also be
allowed for, decoupling reproduction from chemotaxis, etc. We
stress that the avenues explored here, as well as in the
hierarchical model, provide a non-uniform (random) environment for
the growth of towering pillar and mushroom structures. We believe
that these morphotypes are difficult to understand - and indeed do
not emerge easily - in uniform environments assumed when using
standard Eden or DLA type aggregation rules. Incidentally, we
checked that our approach reproduces pure DLA patterns in the
limit of infinite nutrient diffusion (or turbulent convection),
randomizing the nutrient mass over space.

{\bf Acknowledgments.}\\  This research was supported by the
Flemish Programme FWO-G.0222.02 ``Physical and interdisciplinary
applications of novel fractal structures". We thank Katarzyna
Sznajd-Weron  and Jan \.Zebrowski for useful discussions.

\label{}



\end{document}